\documentclass{article}

\usepackage{PRIMEarxiv}

\usepackage[utf8]{inputenc} % allow utf-8 input
\usepackage[T1]{fontenc}    % use 8-bit T1 fonts
\usepackage{hyperref}       % hyperlinks
\usepackage{url}            % simple URL typesetting
\usepackage{booktabs}       % professional-quality tables
\usepackage{amsfonts}       % blackboard math symbols
\usepackage{nicefrac}       % compact symbols for 1/2, etc.
\usepackage{microtype}      % microtypography
\usepackage{lipsum}
\usepackage{xcolor} % Required for color definitions like {green}
\usepackage{fancyhdr}       % header
\usepackage{graphicx}       % graphics
\graphicspath{{images/}} %Setting the graphicspath
\definecolor{mygray}{gray}{0.80}  % 0 = black, 1 = white (so 0.85 is light gray)

\usepackage{amssymb}

\usepackage{placeins}
\usepackage[normalem]{ulem}
\usepackage{multirow}
\usepackage{bbding}
\usepackage[nointegrals]{wasysym}
\usepackage{booktabs}
\usepackage{tabularx}
\usepackage{xltabular}
\usepackage{soul}
\usepackage{array}
\usepackage{adjustbox}
\usepackage{amsmath}
\usepackage{rotating}
\usepackage{longtable}
\usepackage{pdflscape}
\usepackage{threeparttable, caption}
\usepackage{pifont}% http://ctan.org/pkg/pifont

\usepackage{lscape}
\usepackage{xcolor}
\usepackage{caption}
\usepackage{subcaption}
\usepackage{url}
\usepackage{comment}
\usepackage{graphicx}
\usepackage{epstopdf}     % graphics
\usepackage{placeins} 
\usepackage{hyperref}       % hyperlinks
\graphicspath{{images/}} %Setting the graphicspath
\usepackage{url}
\usepackage{comment}
\usepackage{amsmath} 
\usepackage{multirow}
\usepackage[nointegrals]{wasysym}% for \Circle
\usepackage{url}
\usepackage{longtable}
\usepackage{soul}% for hl command 
\usepackage{caption}
\usepackage{subcaption}
\usepackage{ragged2e}
\emergencystretch=\maxdimen
\usepackage{tikz}
\usetikzlibrary{positioning,chains}
\tikzset{
    mynode/.style={
        draw, rectangle, align=center, text width=5cm, inner sep=3ex},
    mylabel/.style={
        draw, rectangle, align=center, rounded corners, font=\bf, inner sep=2ex, 
        fill=cyan!30, minimum height=3.8cm},
    arrow/.style={
        very thick,->,>=stealth}
}

\usepackage{tikz}
\usetikzlibrary{positioning,chains}
\tikzset{
    mynode/.style={
        draw, rectangle, align=center, text width=5cm, inner sep=3ex},
    mylabel/.style={
        draw, rectangle, align=center, rounded corners, font=\bf, inner sep=2ex, 
        fill=cyan!30, minimum height=3.8cm},
    arrow/.style={
        very thick,->,>=stealth}
}

\newcolumntype{R}[2]{%
    >{\adjustbox{angle=#1,lap=\width-(#2)}\bgroup}%
    l%
    <{\egroup}%
}
\title{Enhancing Android Malware Detection with Retrieval-Augmented Generation}

\author{
Saraga S. \\
Department of Computer Applications,\\
Cochin University of Science\\
and Technology, Kochi, India\\
\texttt{saraga@pg.cusat.ac.in} \\
\And
Anagha M. S. \\
Department of Computer Applications,\\
Cochin University of Science\\
and Technology, Kochi, India\\
\texttt{anaghasunil1@pg.cusat.ac.in}\\
\And
Dincy R. Arikkat \\
Department of Computer Applications,\\
Cochin University of Science\\
and Technology, Kochi, India\\
\texttt{dincyrarikkat@cusat.ac.in} \\
\And
Rafidha Rehiman K. A. \\
Department of Computer Applications,\\
Cochin University of Science\\
and Technology, Kochi, India \\
\texttt{rafidharehimanka@cusat.ac.in}
\And
Serena Nicolazzo \\
Department of Computer Science, \\
University of Milan, \\
G. Celoria, 20, Milan, Italy\\
\texttt{serena.nicolazzo@unimi.it} \\
\And
Antonino Nocera \\
Department of Electrical, Computer \\
and Biomedical Engineering, \\
University of Pavia, \\
A. Ferrata, 5, Pavia, Italy \\
\texttt{antonino.nocera@unipv.it} \\
\And
Vinod P. \\
Department of Computer Applications, \\
Cochin University of Science \\
and Technology, Kochi, India \\
\texttt{vinod.p@cusat.ac.in} \\
}

\begin{document}
\maketitle

\begin{abstract}
The widespread use of Android applications has made them a prime target for cyberattacks, significantly increasing the risk of malware that threatens user privacy, security, and device functionality. Effective malware detection is thus critical, with static analysis, dynamic analysis, and Machine Learning being widely used approaches. In this work, we focus on a Machine Learning-based method utilizing static features. We first compiled a dataset of benign and malicious APKs and performed static analysis to extract features such as code structure, permissions, and manifest file content, without executing the apps. Instead of relying solely on raw static features, our system uses an LLM to generate high-level functional descriptions of APKs. To mitigate hallucinations, which are a known vulnerability of LLM, we integrated Retrieval-Augmented Generation (RAG), enabling the LLM to ground its output in relevant context. Using carefully designed prompts, we guide the LLM to produce coherent function summaries, which are then analyzed using a transformer-based model, improving detection accuracy over conventional feature-based methods for malware detection.  
\end{abstract}


\section{Introduction}
The explosive growth of mobile technology has fundamentally transformed the digital landscape, with Android emerging as the dominant operating system for smart devices worldwide. Since its release in 2008, Android has captured an overwhelming market share, accounting for approximately $86.6\%$ of global smartphone sales by 2019. This widespread adoption can be attributed to Android's open architecture and flexibility, characteristics that have simultaneously made it the primary target for sophisticated cyber threats \cite{arzt2014flowdroid}. The Android ecosystem faces unique security challenges stemming from its open-source design and relatively permissive distribution channels for applications, creating vulnerabilities that malicious actors promptly exploit to propagate harmful applications \cite{felt2011android}.
Android malware presents significant threats to users, including unauthorized data exfiltration, financial fraud, and privacy violations, often operating covertly without explicit user consent \cite{zhou2012dissecting}. Although traditional signature-based detection mechanisms effectively identify known threats, they frequently fail to recognize emerging malware variants that employ advanced obfuscation techniques or exploit previously unknown attack vectors \cite{arp2014drebin}. This limitation has prompted researchers to develop more sophisticated approaches, including behavioral analysis, heuristic-based detection, and model checking methodologies, frequently augmented by data mining and Machine Learning algorithms to enhance detection capabilities \cite{tam2015copperdroid}.
The rapidly evolving nature of modern Android malware necessitates a paradigm shift towards more adaptive and complex detection techniques capable of identifying threats without prior exposure. Recent advances in Natural Language Processing (NLP), particularly through Large Language Models (LLMs), offer promising new approaches to addressing these challenges. LLMs have demonstrated exceptional performance in various domains, including text classification and anomaly detection, suggesting significant potential to improve malware classification accuracy \cite{meng2018droidecho}. The integration of NLP capabilities within Deep Learning architectures enables security systems to identify subtle patterns and anomalies in application code and behavior that might otherwise evade traditional detection methods \cite{demontis2017yes}.

This research proposes a system based on AgenticRAG, an emerging approach that combines Retrieval-Augmented Generation (RAG) with agentic behavior specifically tailored for cybersecurity applications, which represents a significant advancement in automated threat detection \cite{zhang2025agentic}. The primary objective is to establish a reliable classification framework that distinguishes between benign and malicious Android applications by analyzing static behavior patterns through advanced NLP techniques. Following application ingestion, the framework performs comprehensive static analysis to extract relevant features characterizing application behavior without requiring execution in a controlled environment \cite{suarez2017droidsieve}. This non-dynamic approach offers considerable advantages in efficiency and scalability when processing large application volumes.
Our innovative methodology leverages advanced NLP techniques to analyze complex technical features of applications while preserving critical security-related properties. The specialized model incorporates domain-specific knowledge and terminology related to malware behavior, allowing a more accurate interpretation of security-related patterns \cite{demontis2017yes}. This approach offers significant advantages, namely, it demonstrates {\em (i)} enhanced detection capabilities through sophisticated pattern recognition and {\em (ii)} improved resistance to evasion techniques commonly employed by malware developers \cite{meng2018droidecho}. Rather than focusing on generating human-readable explanations, our system prioritizes detection accuracy and resilience against adversarial techniques.\\
The multimodal nature of our analysis creates additional barriers that malicious applications must overcome to avoid detection, considering both code properties and natural language descriptions. This adaptability is particularly valuable in the Android ecosystem, where new applications and malware variants emerge daily, creating a constantly evolving security landscape that challenges traditional static approaches \cite{onwuzurike2019mamadroid}. By leveraging the contextual understanding capabilities of transformer-based models, the system can identify subtle indicators of malicious intent that might be overlooked by strictly signature-based approaches, contributing to more robust security for the Android ecosystem.\\
The implications of this study extend beyond immediate practical applications in malware detection. The successful integration of NLP techniques with cybersecurity analytics represents a promising direction for broader digital security applications, including network traffic analysis, vulnerability assessment, and threat intelligence. By bridging the gap between natural language understanding and technical security analysis, this approach opens new possibilities for developing more intuitive and effective security measures that better communicate with human analysts and administrators. The system enables security experts to validate results and incorporate them into broader security strategies, maintaining a crucial human-in-the-loop approach while ensuring AI-based security systems function as effective augmentations of human expertise rather than opaque black boxes.\\
A detailed experimental campaign demonstrates the improvement in the performance of our approach with respect to the state-of-the-art.

In summary, our key contributions, which are part of the proposed system, are the following: 

\begin{itemize}
\item We compile a comprehensive dataset of benign and malicious Android APKs and perform static analysis to extract features without requiring app execution.
\item We propose a novel Android malware detection system that combines Retrieval-Augmented Generation (RAG) with agentic reasoning tailored for cybersecurity, to semantically interpret extracted static features and reveal behavioral patterns, enhancing malware classification accuracy.
\item We perform an extensive experimental evaluation, demonstrating that the proposed system outperforms state-of-the-art baselines in detection accuracy.
\end{itemize}

The remainder of this article follows this structure: Section~\ref{sec:related_work} discusses the current literature related to our proposal. Section~\ref{sec:design} provides a detailed explanation of our framework and its architecture. We discuss the experiment results and evaluation in Section~\ref{sec:result}. Finally, Section~\ref{sec:conclusion} concludes the paper.

\section{Related Work}
\label{sec:related_work}
Android malware detection has evolved from signature-based methods to sophisticated AI approaches. Early techniques identified known malware at $87-92\%$ accuracy but struggled with polymorphic malware \cite{ rastogi2013droidchameleon}, while the effectiveness dropped below $60\%$ against zero-day attacks \cite{ suarez2013evolution}. Dynamic analysis techniques emerged in response, with pioneering system-call monitoring approaches achieving $93\%$ detection accuracy by observing runtime behaviors \cite{yan2012droidscope}.
Deep Learning transformed the field with the introduction of CNNs, which converts binary code into image representations, achieving $96\%$ detection accuracy while reducing false positives by $40\%$ \cite{ wang2019effective}.  Building on this success, advanced the field using Graph Neural Networks that model relationships between application components, achieving $98.3\%$ detection accuracy on benchmark datasets with remarkable resilience against adversarial evasion techniques, maintaining above $95\%$ accuracy even against disguised malicious code \cite{hou2017hindroid}. This evolution mirrors trends regarding how LLMs have revolutionized research approaches through increasingly sophisticated pattern recognition capabilities \cite{ liu2023summary}.\\
A significant breakthrough came with AppPoet \cite{ zhao2025apppoet}, which pioneered multi-view prompt engineering for Android malware detection. By analyzing applications from multiple perspectives simultaneously (code structure, permissions, API usage), AppPoet achieved $99.3\%$ detection accuracy while demonstrating exceptional resilience to obfuscation (only $3.2\%$ in performance degradation). Its novel prompt engineering methodology dynamically adapts to emerging malware characteristics, representing a shift from static feature engineering to adaptive, context-aware analysis. AppPoet also demonstrated enhanced robustness against adversarial examples, maintaining over $90\%$ in detection accuracy even when subjected to state-of-the-art evasion techniques.

Despite these advances, critical challenges remain open. Transformer-based systems exhibit significant position dependent bias \cite{zheng2023judging}. It revealed consistency rates as low as $11.2\%$ to $16.2\%$ in zero-shot settings, with some models showing up to $75\%$ bias toward the first position. Even advanced models like GPT-4 showed position bias, though it improved from $65.0\%$ to $77.5\%$ in consistency with few-shot prompting. Furthermore, these systems are susceptible to verbosity effects, where unnecessarily verbose responses are artificially rated higher, a phenomenon which was demonstrated through their ``repetitive list'' attack that successfully misled multiple LLM judges \cite{zheng2023judging}. AppPoet addressed these concerns through controlled experimental design, establishing new standards for robust evaluation. However, computational demands remain significant, with inference times exceeding 5 seconds per application on resource-constrained devices, a limitation for the LLM-based approach in \cite{ zhao2025apppoet}.

This review highlights that, while static analysis provides an efficient foundation for Android malware detection and feature engineering is crucial for optimizing input data, traditional text-based Machine Learning approaches face significant hurdles due to malware evolution, obfuscation, and inherent model limitations. Advanced models, particularly BERT, demonstrate considerable promise by generating robust feature representations \cite{sun2024research}, exhibiting greater resilience to adversarial attacks \cite{zhao2025apppoet}. Therefore, investigating and implementing sophisticated BERT-based classification architectures presents a compelling direction for advancing the state-of-the-art in accurately and robustly differentiating malicious Android applications from benign ones, while addressing the computational and evaluation limitations of prior methods. This constitutes the primary research objective undertaken in this study.

\section{System Architecture and Methodology}
\label{sec:design}
\begin{figure*}
    \centering
    \includegraphics[width=1\textwidth]{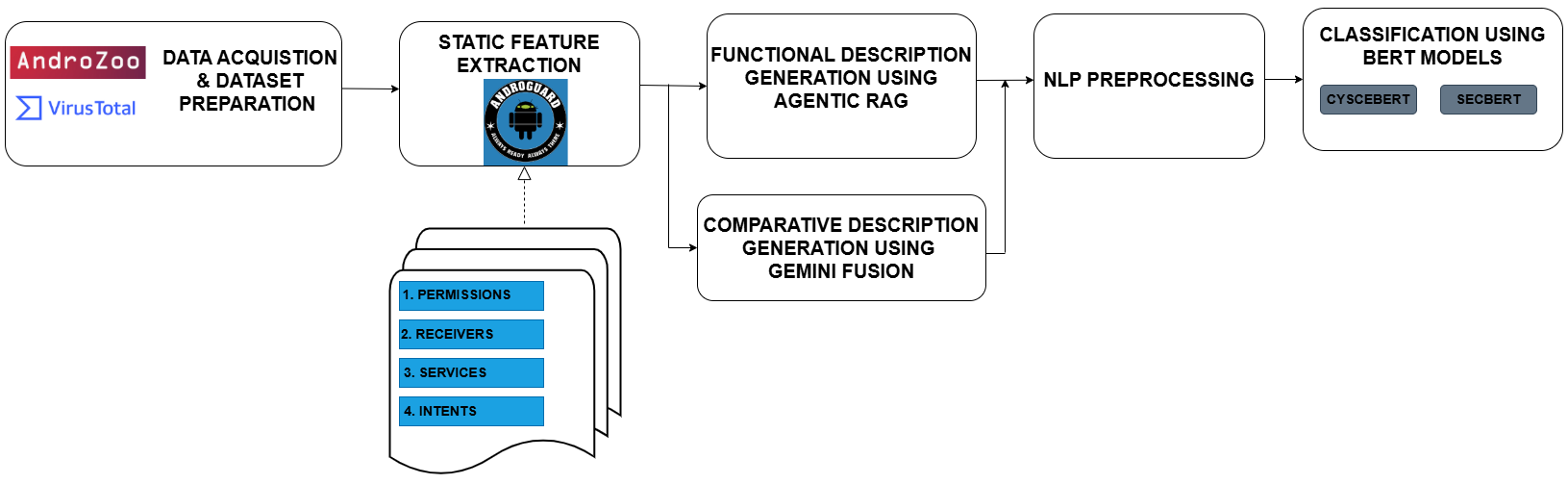}
    \caption{Architecture of the proposed Android Malware Classification System incorporating APK based Feature Extraction and Functional Description Generation using AgenticRAG }
    \label{fig:Methodology}
\end{figure*}

The following section describes the procedures used for data acquisition and the methodological framework applied in this research. 
As shown in Figure \ref{fig:Methodology}, the proposed system for Android malware analysis consists of the following main components:
\begin{enumerate}
    \item \textbf{Dataset Acquisition and Dataset Preparation}. We collected a comprehensive dataset of Android applications, including both benign and malicious APK files, from publicly available sources.
    \item \textbf{Static Feature Extraction}. Using static analysis techniques, we extracted key features from the APKs without executing them.
    \item \textbf{Functional Description Generation}. Extracted features were used as input for a Large Language Model (LLM) to generate high-level functional descriptions of each application. Two approaches were used, namely {\em (i)} AgenticRAG approach and {\em (ii)} Gemini Fusion approach.
    \item \textbf{NLP Preprocessing}. The generated descriptions underwent Natural Language Preprocessing to prepare the data for downstream classification.
    \item \textbf{Text Classification}. Two transformer-based text classification models based on BERT were trained on the preprocessed functional descriptions to distinguish between benign and malicious applications.
\end{enumerate}

In the following subsections, we describe in detail all the steps of our framework. Table \ref{tab:SystemSymbols} summarizes the most employed acronyms used throughout the paper.

\begin{table}
\centering
\caption{Summary of the acronyms used in the paper}
\label{tab:SystemSymbols}
\begin{tabular*}{\columnwidth}{@{\extracolsep{\fill}}ll@{}}
\hline
\textbf{Symbol} & \textbf{Description}\\
\hline
APK & Android Package Kit \\
    BERT & Bidirectional Encoder Representations \\
    & from Transformers \\
    CNN & Convolutional Neural Network \\
    DL & Deep Learning \\
    LLM & Large Language Model \\
    ML & Machine Learning \\
    NLP & Natural Language Processing \\
    RAG & Retrieval-Augmented Generation \\
    \hline
\end{tabular*}
\end{table}

\subsection{Data Acquisition and Dataset Preparation}

The corpus of Android applications utilized in this investigation was systematically obtained from AndroZoo\footnote{\url{https://orbilu.uni.lu/bitstream/10993/27396/1/androzoo.pdf}}, an extensive repository that houses millions of Android Package Kit (APK) files aggregated from various distribution channels. Each specimen within the AndroZoo collection is accompanied by comprehensive metadata, including cryptographic identifiers in the form of SHA256 hash values, which facilitate cross-referencing with external threat intelligence infrastructures.\\
For the purpose of establishing a reliable classification taxonomy that distinguishes between benign and malicious applications, this research leveraged the analytical capabilities of the VirusTotal\footnote{\url{https://www.virustotal.com/gui/home/searchhttps://www.virustotal.com/gui/home/search}} application programming interface. VirusTotal constitutes a sophisticated security ecosystem that synthesizes detection outcomes from over sixty distinct antivirus engines, delivering comprehensive security assessments in structured JavaScript Object Notation (JSON) format.\\
The methodological framework for classification label assignment adhered to a systematic protocol. Initially, each application's unique SHA256 cryptographic hash value was utilized to retrieve its corresponding VirusTotal JSON analytical report. These reports encapsulated comprehensive detection results from multiple security engines, providing a multidimensional perspective on potential malicious attributes.\\
Subsequently, a threshold-based heuristic criterion was implemented for binary classification purposes. Applications that activated detection mechanisms in one or more security engines were classified as malicious entities, reflecting the conservative approach prioritized in security contexts. In contrast, applications that evaded detection in all integrated antivirus platforms were classified as benign applications. This binary classification schema established the foundational ground truth for subsequent analytical procedures in this investigation.

\subsection{Static Feature Extraction}
After data collection, we implemented static analysis methodologies. They were developed through the application of Androguard\footnote{\url{https://github.com/androguard/androguard}}, a sophisticated reverse engineering framework specifically designed for the deconstruction and analysis of Android applications. This analytical approach facilitates the extraction of characteristic features from application packages without necessitating their execution in runtime environments, thereby providing substantial insights into architectural composition and behavioral propensities.\\
The extraction protocol focused on several categories of static features with significant security implications:\\
\begin{itemize}
    \item \textbf{Permissions} embedded within the AndroidManifest.xml configuration file were systematically identified and cataloged. These permissions constitute formal requests for access to protected system resources and functionalities, such as geolocation data, camera hardware, Short Message Service (SMS) capabilities, and other sensitive device interfaces. The pattern of requested permissions provides crucial indicators regarding an application's operational scope and potential security implications.
    \item \textbf{Receivers} implemented within the application were identified and examined. These components enable the application to respond to system-wide broadcast events and messages, establishing pathways through which the application interacts with system notifications and external stimuli. The configuration of these receivers illuminates significant dimensions of an application's responsiveness to environmental conditions.
    \item \textbf{Services} defined within the application were comprehensively analyzed. These components perform background operations without direct user interaction and can execute even when the application is not in the foreground, potentially enabling persistent operational capabilities. The identification and characterization of these services reveal critical aspects of an application's background processing model and long-running functionalities.
    \item \textbf{Intents} that are action declarations associated with broadcast receivers and service components were extracted and analyzed. These action declarations specify the programmatic behaviors triggered in response to system events or custom notifications, providing insight into an application's reactive mechanisms and event handling protocols.
\end{itemize}

This constellation of static features collectively constitutes a comprehensive representation of the application's architectural design, operational capabilities, and potential security implications. The feature set captures both legitimate functional requirements and potential indicators of malicious intent, such as excessive permission requests or suspicious component configurations. These extracted characteristics serve as fundamental determinants of an application's security profile, enabling subsequent analytical processes to identify patterns consistent with either benign functionality or malicious exploitation of system resources.

\subsection{Functional Description Generation using AgenticRAG}

Although static features such as permissions and intent filters provide significant insights into application behavior, these attributes represent low-level technical constructs that frequently present interpretability challenges for both human analysts and NLP models. To transform these granular static features into semantically rich and coherent textual representations, this research implemented a custom AgenticRAG (Agentic Retrieval-Augmented Generation) system built using Gemini 2.0 Flash Lite. This sophisticated hybrid architecture seamlessly integrates information retrieval mechanisms with generative artificial intelligence capabilities to synthesize comprehensive behavioral summaries of Android applications automatically.

The AgenticRAG framework addresses the fundamental challenge of translating technical application characteristics into natural language narratives that capture the essential behavioral patterns and security implications while maintaining fidelity to the underlying technical attributes. By bridging the semantic gap between raw feature vectors and interpretable descriptions, this approach facilitates more effective utilization of transformer-based models for security classification tasks.\\
The foundation of the AgenticRAG system rests upon a comprehensive Data Corpus that was meticulously constructed through systematic extraction using custom web crawlers and consolidation of authoritative information from the official Android Developer documentation\footnote{\url{https://developer.android.com}}. The Data Corpus comprises four distinct Excel sheets, each corresponding to a specific category of static features, with each database entry structured to contain the precise technical feature name as referenced in Android APIs alongside comprehensive human-readable descriptions elucidating the feature's intended purpose and behavioral characteristics.
The Data Corpus operates on four distinct categories of static features:

\begin{itemize}
    \item {\em Permissions}: Authorization controls that determine application access rights to protected system resources and user data.
    \item {\em Services}: Background tasks that run without a UI. Reveals hidden operations like data uploads or downloads.
    \item {\em Receivers}: Components designed to intercept and process broadcast messages and system events.
    \item {\em Intent Actions}: Intent-based operations that trigger specific behaviors and functionality across application boundaries.
\end{itemize}
This resulting database serves as the backbone of the ensemble retrieval module.\\

\begin{figure*}
    \centering
    \includegraphics[width=\textwidth]{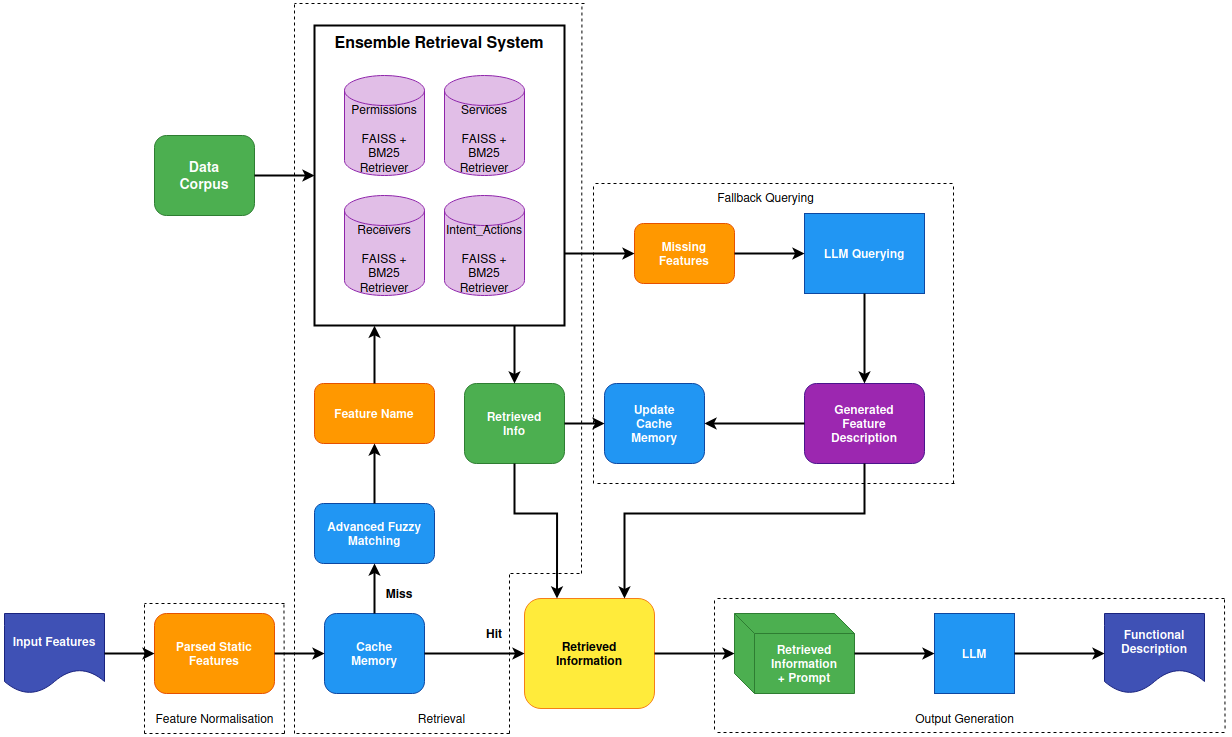}
    \caption{Architecture of the AgenticRAG System}
    \label{fig:AgenticRAG}
\end{figure*}

Our AgenticRAG system combines the capabilities of structured retrieval and large language model reasoning within a goal-driven, memory-enabled agent architecture. A focus on this component is illustrated in Figure \ref{fig:AgenticRAG}, where different steps are shown, such as {\em (i)} feature normalization, {\em (ii)} retrieval, {\em (iii)} fallback querying, and {\em (iv)} output generation through a structured pipeline that begins with input features and culminates in functional descriptions.

\textbf{Feature Normalization}. This refers to the structured extraction and normalization of raw AndroidManifest attributes such as permissions, services, receivers, and intent actions. This involves cleaning inconsistent formatting, converting technical identifiers to standardized forms, and mapping each feature to its appropriate category. The resulting parsed static features serve as inputs to the retrieval and reasoning components, enabling reliable and semantically aligned processing across the pipeline.

\textbf{Retrieval}. The system first consults its Cache Memory to determine if descriptions for the features have been previously retrieved or generated. In case of a cache hit, the system directly accesses the Retrieved Information. When features are not found in cache (cache miss), the system routes them to the Ensemble Retrieval System for processing.
Each feature category from the Data Corpus is supported by a dedicated FAISS + BM25 Retriever combination. The retrieval module utilizes an ensemble retriever that combines two complementary retrieval mechanisms: {\em (i)} a FAISS-based dense vector retriever and {\em (ii)} a BM25-based sparse keyword retriever. The FAISS retriever is powered by HuggingFace embeddings and captures semantic similarity, while the BM25 retriever focuses on exact keyword overlap. These are integrated using the Reciprocal Rank Fusion (RRF) technique used by the Ensemble retriever, with configurable weights allowing the system to prioritize semantic or lexical matches.
When a static feature is encountered, it undergoes the Advanced Fuzzy Matching process. To ensure robust feature identification across varied static inputs, the system implements this two-stage matching mechanism. Upon receiving a normalized feature string, the system first attempts an exact match against the relevant dataset from the Data Corpus. If a direct correspondence is found, it is immediately accepted as a high-confidence match and processed by the appropriate ensemble retriever. In the absence of an exact match, the system proceeds to a secondary fuzzy matching stage, wherein the input is compared against all available keys using a string similarity metric based on Levenshtein distance. The best approximate match is selected if its similarity score exceeds a configurable threshold (e.g., $65\%$). This two-stage strategy—prioritizing exact match for precision and applying fuzzy match for resilience—effectively captures partial, truncated, or misspelled feature names, thereby improving recall without compromising correctness in retrieval. The ensemble retriever then outputs the Retrieved Feature Description.

\textbf{Fallback querying}. For Missing Features that cannot be matched through either exact or fuzzy matching (falling below the similarity threshold), the system groups them together as a batch and queries the LLM (Gemini 2.0 Flash Lite). The LLM generates feature descriptions, which are then added to the Retrieved Information and also stored in the Cache Memory component to enhance future retrieval efficiency.

\textbf{Output generation}. Following successful retrieval or generation, the Retrieved Information is combined with a structured prompt, which is shown in Table \ref{tab:prompt1} and passed to the LLM component. The LLM synthesizes this information to produce the final Functional Description, completing the pipeline.

\begin{table*}
  \centering
  \caption{Prompt for description generation using AgenticRAG}
  \label{tab:prompt1}
  \fcolorbox{black}{mygray}{\parbox{\linewidth}{
\ttfamily
\large
\textbf{Role :} 
You are an Android security expert analyzing an app named \textbf{\{apk\_name}\}. Your task is to generate a concise, factual summary of the app's functionality strictly based on its static features.
\vspace{1em}

\textbf{Input Information:} App Statistics:

- Permissions: \{feature\_stats[`total\_permissions']\}

- Services: \{feature\_stats[`total\_services']\}

- Broadcast Receivers: \{feature\_stats[`total\_receivers']\}

- Intent Actions: \{feature\_stats[`total\_intents']\}
\vspace{1em}

Here is the technical information about the app's features: \textbf{\{formatted\_info\}}
\vspace{1em}

\textbf{Response Format:} 
Based on these features, write a single cohesive paragraph that:

1.  Describes what the app can do and its main functionality.

2.  Notes any significant security-relevant capabilities.

3.  Explains how the different features work together.

4.  Is factual and objective without speculation.

5.  Does not provide any additional comments or assumptions.

6.  Does not suggest further analysis.

7.  Does not include \textbf{\{apk\_name\}} in the final description.

\vspace{1em}
Focus only on what the app CAN do, NOT what it MIGHT do. Be specific about functionality, not generic.
\\
}
}
\end{table*}

The generation process proceeds through a systematic sequence of operations:
\begin{enumerate}
    \item The feature set is transformed into a pseudo-natural language query, converting technical specifications into a more accessible format for the language model.
    \item Retrieved descriptions from the ensemble retrieval mechanism are incorporated as comprehensive background context, providing the model with domain-specific knowledge about Android features and their functionalities.
    \item The LLM is explicitly instructed to generate a coherent functional description of the application, synthesizing the technical features with their practical implications.
\end{enumerate}

The overall analysis process iterates over each APK's extracted static features, applies the full pipeline shown in Figure \ref{fig:AgenticRAG}, and logs the resulting functional descriptions. The system exhibits agentic properties by autonomously managing the end-to-end decision-making process involved in Android app analysis. It decomposes its task into modular reasoning steps, dynamically selects between structured retrieval and generative fallback based on context, and leverages persistent memory through the Cache Memory component to optimize future responses. This enables the system to operate with autonomy, adaptability, and goal-driven behavior—core characteristics of an agentic architecture reflected in the comprehensive pipeline architecture. The resultant textual outputs underwent NLP preprocessing routines to standardize formatting and enhance semantic clarity. These refined descriptions subsequently served as the primary input to the BERT classifier for further analysis and categorization. This approach leverages the generative capabilities of Gemini 2.0 Flash Lite while constraining its output through structured prompting and contextual grounding, ensuring technically accurate yet accessible descriptions of application functionality.

\subsubsection{Comparative Description Generation Approaches}
\begin{figure}
    \centering
    \includegraphics[width=0.7\textwidth]{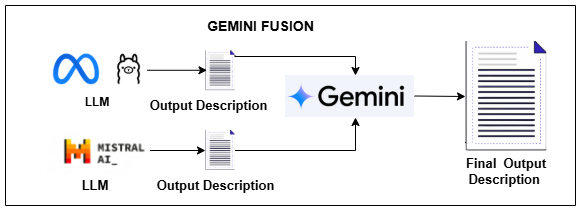} 
    \caption{Architecture of Gemini Fusion}
    \label{fig:Gemini}
\end{figure}

The research methodology incorporated a parallel description generation approach using a sophisticated model fusion strategy. In addition to the AgenticRAG pipeline, a comparative framework was established utilizing Google's Gemini model in conjunction with open-source LLMs depicted in Figure \ref{fig:Gemini}. We refer to it as the Gemini fusion methodology.\\
Using Gemini 2.0 Flash Lite via API is preferred due to its speed, stability, and ease of integration, requiring no local hardware or complex setup. While using other models like LLaMA and Mistral, quantization becomes necessary to reduce memory and computational demands, which often leads to a loss in accuracy and increased engineering overhead. Hence, Gemini provides a more efficient and reliable solution for lightweight, high-performance NLP tasks.

The Gemini Fusion methodology employs a multi-stage process:
\begin{itemize}
    \item Independent description generation was performed using two prominent open-source LLMs {\em(i)} LLaMA2 (Large Language Model Meta AI) and {\em(ii)} Mistral.
    \item These independently generated descriptions were subsequently integrated through Gemini 2.0 Flash Lite's instruction-following capabilities, creating a synthesized output that leveraged the semantic strengths of each constituent model.
    \item This approach effectively simulates collaborative reasoning among diverse model architectures, potentially capturing complementary aspects of application functionality.
\end{itemize}
Table \ref{tab:prompt21} and Table \ref{tab:prompt22} are the prompts used for the purpose.

\begin{table*}
  \centering
  \caption{Prompt for description generation using Llama and Mistral}
  \label{tab:prompt21}
  \fcolorbox{black}{mygray}{\parbox{\linewidth}{
\ttfamily
\large
\textbf{Role :} 
You are an Android security expert. Your task is to generate a concise, factual summary of the app's functionality strictly based on its static features.
\vspace{1em}

\textbf{Input Information:}\\
<INPUT>
\{formatted\_info\}
</INPUT>
\vspace{1em}

\textbf{Response Format:} 
Based on the Android app information above, write a single paragraph describing the app's functionality without any additional comments or assumptions.

\vspace{1em}
}
}
\end{table*}

\begin{table*}
  \centering
  \caption{Prompt for description generation using Gemini for fusion}
  \label{tab:prompt22}
  \fcolorbox{black}{mygray}{\parbox{\linewidth}{
\ttfamily
\large
\textbf{Role :} 
You are an expert in the field of Android security, specializing in auditing Android applications by static analysis.
\vspace{1em}

\textbf{Input Information:}

Below are the details for an APK:\\
APK Name: \{apk\_name\}\\
Static Features:\\
\{static\_features\_str\}\\
Description 1: \{description1\}\\
Description 2: \{description2\}
\vspace{1em}

Your task is to generate a new functional description for the APK by analysing the \{static\_features\_str\}. You also have to incorporate the best aspects of both \{description1\}  and \{description2\} if applicable in your final description.
\vspace{1em}

\textbf{Response Format:} \\
**Instructions:**\\
1. Do not include any comments or comparisons.\\
2. Provide only the final description.\\

**Final Description:**

\vspace{1em}
}
}
\end{table*}

The research design established a controlled comparative framework. Each Android application in the dataset was associated with dual functional descriptions, namely {\em(i)} AgenticRAG-generated description, and {\em(ii)} Gemini Fusion-generated description. Both description sets underwent identical NLP procedures to ensure methodological consistency.
The preprocessed descriptions from both generation strategies will be classified using the same model. Performance metrics will be calculated independently for each description generation approach to facilitate direct comparison of their effectiveness in supporting malware classification.

This dual-pipeline methodology enables rigorous evaluation of how different description generation strategies impact downstream classification performance, providing insight into the relative merits of AgenticRAG versus model fusion approaches for security-oriented application analysis.

\subsection{NLP Preprocessing of Functional Descriptions}
Prior to submitting the functional descriptions to the classification model, comprehensive NLP preprocessing was implemented to ensure textual consistency and semantic clarity. This methodical preprocessing phase enhances both the performance metrics and generalization capabilities of the BERT model, addressing critical challenges in natural language understanding.
\begin{itemize}
    \item \textbf{Text Cleaning: }All generated descriptions underwent systematic purification, removing superfluous punctuation, extraneous special characters, and redundant spacing elements. This sanitization process creates a standardized textual foundation, eliminating potential sources of noise that could otherwise confound the classification algorithm. Inconsistent character patterns or formatting anomalies might otherwise be misinterpreted as semantically meaningful features.
    \item \textbf{Lowercasing: }The entire textual corpus was transformed to lowercase format, effectively eliminating case sensitivity variables that might otherwise introduce inconsistencies in the classification process. This normalization is particularly important when working with pre-trained language models like BERT, which may associate different embeddings with identical words presented in varying cases.
    \item \textbf{Stopword Removal: }Common English stopwords (such as ``is'', ``the'', ``of'') were systematically identified and extracted from the text, as these linguistic elements contribute minimal semantic value while potentially introducing noise into the feature representation. By removing these high-frequency but low-information words, the system can focus computational resources on the truly distinctive and meaningful components of app descriptions.
    \item \textbf{Stemming: }Porter stemming algorithms were applied to reduce lexical variations to their root forms. For example, morphological variants like "sending," "sends," and "sent" were all consolidated to the root form "send." This normalization technique substantially reduces representational redundancy in the feature space, enabling the model to recognize functional equivalence despite surface-level morphological differences.
\end{itemize}
The necessity of this preprocessing pipeline stems from the inherent challenges of Natural Language Processing. Unprocessed text contains numerous variations and inconsistencies that can obscure underlying semantic patterns. By standardizing the input format, reducing dimensionality, and eliminating noise, preprocessing creates an optimal feature space for cybersecurity-specific BERT variants such as CySecBERT and SecBERT. This enhances the model's ability to discern meaningful patterns rather than focusing on superficial textual variations, ultimately improving classification accuracy and robustness.\\
Furthermore, this preprocessing approach ensures that descriptions generated by different iterations of the AgenticRAG system maintain consistent formatting, creating a reliable pipeline from generation through classification. The result is a system capable of transforming diverse functional descriptions into standardized representations that maximize the effectiveness of the subsequent classification operations.

\subsection{Classification Using BERT Models}

The choice of BERT-based models for Android application classification is motivated by several practical advantages over larger language models. Smaller models like BERT offer faster inference times, lower computational requirements, and reduced memory footprint, making them ideal for deployment in resource-constrained environments or real-time classification scenarios. Additionally, these models can be fine-tuned effectively with smaller datasets while maintaining good generalization performance, which is particularly valuable in cybersecurity applications where labeled data may be limited.

Textual data is tokenized using the model's native tokenizer, and a custom dataset class is implemented to facilitate seamless integration with the HuggingFace Trainer API. The classification of Android applications is conducted using CySecBERT \cite{bayer2024cysecbert} and SecBERT \cite{huang2024secbert}, specialized variants of the BERT architecture that have been specifically optimized for cybersecurity contexts. These domain-specific models leverage pre-training on cybersecurity corpora, enabling them to better understand security-related terminology and concepts compared to general-purpose language models. The models undergo fine-tuning on preprocessed application descriptions through a supervised learning approach, allowing them to learn the specific patterns and characteristics that distinguish between different categories of Android applications.
%Textual data is tokenized using the model's native tokenizer, and a custom dataset class is implemented to facilitate seamless integration with the HuggingFace Trainer API. The classification of Android applications is conducted utilizing CySecBERT \cite{bayer2024cysecbert} and SecBert \cite{huang2024secbert}, specialized variants of the BERT architecture optimized for cybersecurity contexts. This model undergoes fine-tuning on preprocessed application descriptions through a supervised learning paradigm.

\section{Experimental Campaign}
\label{sec:result}
This section describes the experiments carried out to test the performance of our approach. In particular, in the following sections, we will describe the dataset,
and the evaluation metrics used to test our framework, and we report the obtained results along with their critical analysis.

\subsection{Dataset and Setup}
The corpus comprises a substantial dataset of $10,000$ benign and $8,000$ malicious samples, providing a comprehensive representation of both classes.
Following the assignment of binary classification labels, the dataset is partitioned into training, validation, and testing subsets with a distribution ratio of 70:10:20, while meticulously preserving the class distribution across all partitions. 
To mitigate the effects of class imbalance in the dataset, class weights are calculated and subsequently incorporated in conjunction with a weighted random sampling strategy. The model training process is orchestrated using the Trainer class, with systematic evaluation conducted at the conclusion of each epoch and an early stopping mechanism implemented based on validation loss trajectory to prevent overfitting.

Post-training, a comprehensive evaluation is performed on the segregated test set, with multiple performance metrics computed, including accuracy and F1 score. The finalized model and its corresponding tokenizer are preserved for subsequent inference operations.\\

\begin{comment}
\begin{table}
\addcontentsline{lot}{section}{6.8} 
\centering
\begin{tabular}{|c|l|}
\hline
\textbf{Metrics} & \textbf{Descriptions} \\
\hline
$TP$ & Number of correctly identified malicious Apps \\
$TN$ & Number of correctly identified benign Apps \\
$FP$ & Number of misidentified benign Apps \\
$FN$ & Number of misidentified malicious Apps \\
$Accuracy$ & $(TP + TN) / (TP + TN + FP + FN)$ \\
$Precision$ & $TP / (TP + FP)$ \\
$Recall$ & $TP / (TP + FN)$ \\
$F1-Score$ & $(2 \times Precision \times Recall) / (Precision + Recall)$ \\
\hline
\end{tabular}
\caption{Description of evaluation metrics.}
\end{table}
\end{comment}
\subsection{Evaluation Metrics}
To assess the performance of our framework, we employed the classical ML evaluation metrics, such as Accuracy, Precision, Recall, and F1-score. 
\begin{itemize}
    \item \textit{Accuracy (A)}. It measures the proportion of samples that were correctly classified by the model. 
$$\text{A} = \frac{1}{N} \sum_{i=1}^{N} (y_i = \hat{y}_i)$$
    
    where N is the total number of instances, $y_i$ is the true label for the $i$-th instance, and $\hat{y}_i$ is the predicted label for the $i$-th instance.

    \item \textit{Precision (P)}. This metric evaluates the proportion of correctly predicted labels among all predicted labels and is defined as:

    $$\text{P} = \frac{\text{TP}}{\text{TP} + \text{FP}}$$
    
    where TP (True Positives) refers to the number of instances correctly predicted as belonging to a particular class, or the number of correctly identified malicious applications. FP (False Positives) denotes the number of instances incorrectly predicted as belonging to that class, or the number of misidentified benign applications.

%Weighted precision ($P_{weighted}$) adjusts for class imbalance by computing the precision for each label and weighting it by the number of true instances. Macro precision ($P_{macro}$), on the other hand, calculates the unweighted mean of precision across all labels, treating all classes equally regardless of their frequency.

    \item \textit{Recall (R)}. Recall measures the proportion of correctly predicted labels among all true labels and is defined as:

    $$\text{R} = \frac{\text{TP}}{\text{TP} + \text{FN}}$$
    
    where FN (False Negatives) refers to the number of instances that belong to a class but were not predicted as such. In our case, it is the number of misidentified benign applications.
    
   \item \textit{F1-score (F1)}. This metric is the harmonic mean of precision and recall, and is defined as:

   $$ \text{F1} = 2 \cdot \frac{\text{P} \cdot \text{R}}{\text{P} + \text{R}}$$
   
\end{itemize}

\subsection{Results}
To determine the most effective sentence model for classification, we compared specialized BERT-based models,CySecBERT and SecBERT.\\
To rigorously assess the effectiveness of the AgenticRAG framework in enhancing Android malware classification, we conducted a comprehensive comparative analysis against an alternative description generation approach. The comparative framework leveraged Google's Gemini model to create a fusion of descriptions independently generated by two leading open-source LLMs—LLaMA and Mistral. This fusion strategy was designed to simulate collaborative reasoning between multiple models and served as a competitive baseline.\\
We also conducted a comparative analysis of different LLMs for fusion to determine the most effective LLM. Our analysis compared Gemini 2.0 Flash Lite, Llama2, and Mistral as fusion models.

\subsubsection{Classification Model Performance Comparison}
\begin{figure*}
  \centering
  \addcontentsline{lof}{section}{6.8.2} 
  % First row
  \begin{subfigure}[b]{0.47\textwidth}
    \centering
    \includegraphics[width=\textwidth]{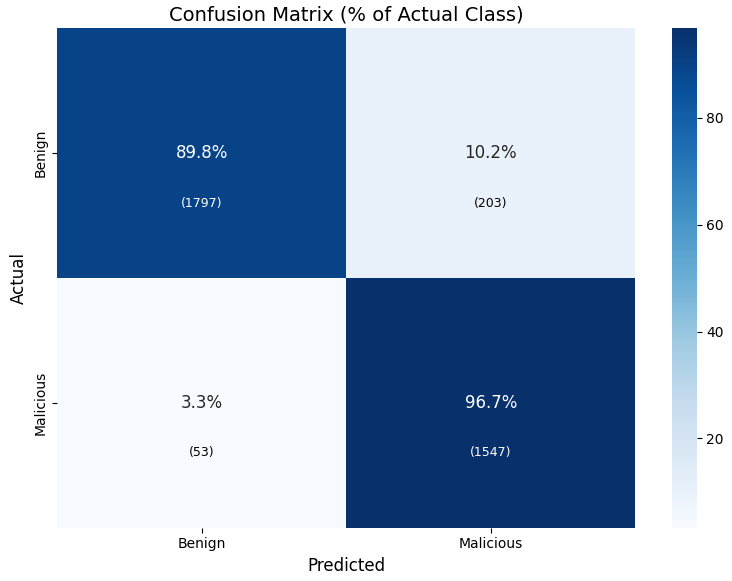}
    \caption{}
  \end{subfigure}
  \hfill
  \begin{subfigure}[b]{0.485\textwidth}
    \centering
    \includegraphics[width=\textwidth]{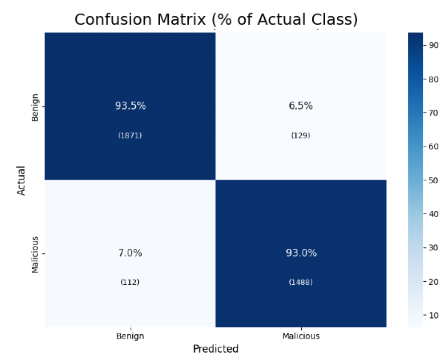}
    \caption{}
  \end{subfigure}

  \caption{Confusion Matrix for (a) AgenticRAG Descriptions Classified using CySecBert, (b)AgenticRAG Descriptions Classified using SecBert}
  \label{fig:CM1}
\end{figure*}

\begin{figure*}
  \centering
  \addcontentsline{lof}{section}{6.8.2} 
  
  \begin{subfigure}[b]{0.47\textwidth}
    \centering
    \includegraphics[width=\textwidth]{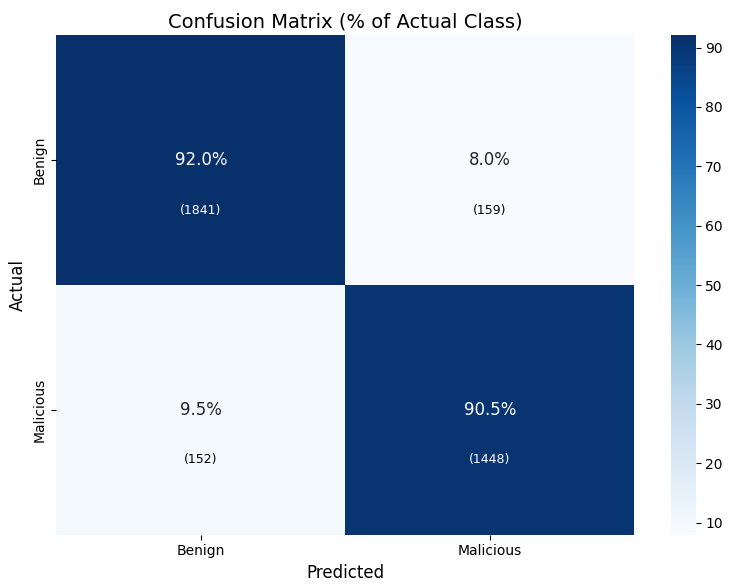}
    \caption{}
  \end{subfigure}
  \hfill
  \begin{subfigure}[b]{0.49\textwidth}
    \centering
    \includegraphics[width=\textwidth]{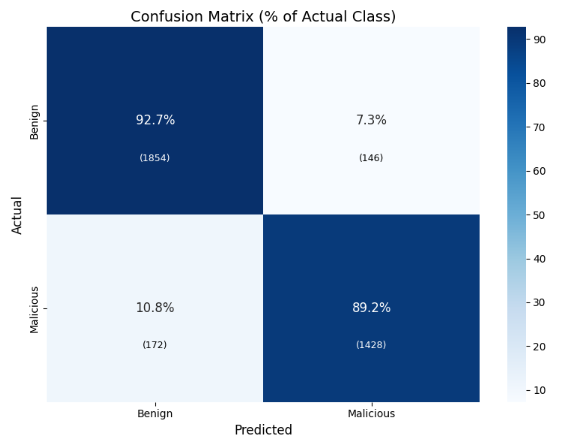}
    \caption{}
  \end{subfigure}

  \caption{Confusion Matrix for (a) Gemini Fusion Descriptions Classified using CySecBert, (b) Gemini Fusion Descriptions Classified using SecBert}
  \label{fig:CM2}
\end{figure*}

This section is devoted to a comparative analysis of specialized BERT-based models to determine the most effective classification architecture. Our analysis compared CySecBERT\footnote{\url{https://huggingface.co/markusbayer/CySecBERT}}, specifically fine-tuned for cybersecurity domain language patterns, against SecBERT\footnote{\url{https://huggingface.co/jackaduma/SecBERT}}, another security-oriented transformer model. The results are shown in Table \ref{tab:model_comparison1} and Table \ref{tab:model_comparison2}. We can see that CySecBERT with a recall rate of $96.89\%$ (see the Confusion Matrix in Figure \ref{fig:CM1}) and an F1-Score of $92.86\%$ for AgenticRAG description and a recall rate of $90.50\%$ (see the Confusion Matrix in Figure \ref{fig:CM2}) and an F1-Score of $91.25\%$ for Gemini Fusion description, outperforms SecBERT in classification.

\begin{table}
\centering
\caption{Performance Comparison of BERT-based classification models for AgenticRAG description}
\label{tab:model_comparison1}
\begin{tabular*}{\columnwidth}{@{\extracolsep{\fill}}lcc@{}}
\hline
 \textbf{Metric} & \textbf{CySecBERT} & \textbf{SecBERT}\\
\hline
    Accuracy & 92.89\% & 93.31\% \\
    Precision & 88.40\% & 92.02\%  \\
    Recall & 96.69\% & 93.00\% \\
    F1-Score & 92.86\% & 92.51\%\\
    \hline
\end{tabular*}
\end{table}

\begin{table}
\centering
\caption{Performance Comparison of BERT-based classification models for Gemini Fusion description}
\label{tab:model_comparison2}
\begin{tabular*}{\columnwidth}{@{\extracolsep{\fill}}lcc@{}}
\hline
 \textbf{Metric} & \textbf{CySecBERT} & \textbf{SecBERT}\\
\hline
    Accuracy & 91.36\% & 91.17\% \\
    Precision & 90.07\% & 90.73\%  \\
    Recall & 90.50\% & 89.25\% \\
    F1-Score & 91.25\% & 91.04\%\\
    \hline
\end{tabular*}
\end{table}

\subsubsection{Description Generation Performance Comparison}
The descriptions considered in our study, namely AgenticRAG-generated and Gemini Fusion-generated, underwent identical NLP preprocessing and classification using the fine-tuned CySecBERT model. The quantitative results shown in Table \ref{tab:description_comparison} demonstrate a clear performance advantage for the AgenticRAG approach, with a recall rate of $96.67\%$, where the Gemini fusion approach only has a recall rate of $90.50\%$. The AgenticRAG approach also has an accuracy of $92.89\%$ and an F1-score of $92.86\%$, outperforming the Gemini fusion approach.

\begin{table}
\centering
\caption{Performance Comparison for Generation Approaches description}
\label{tab:description_comparison}
\begin{tabular*}{\columnwidth}{@{\extracolsep{\fill}}lcc@{}}
\hline
\textbf{Metric} & \textbf{AgenticRAG} & \textbf{Gemini Fusion}\\
\hline
    Accuracy & 92.89\% & 91.36\% \\
    Precision & 88.40\% & 90.07\%  \\
    Recall & 96.69\% & 90.50\% \\
    F1-Score & 92.86\% & 91.25\%\\
    \hline
\end{tabular*}
\end{table}

\begin{table*}
\caption{Performance Comparison of different Fusion models classified using CySecBERT and SecBERT}
\label{tab:model_comparison3}
\begin{tabular*}{\textwidth}{@{\extracolsep{\fill}}llllll@{}}
\hline

    \textbf{Fusion Model} & \textbf{Classifier} & \textbf{Accuracy} & \textbf{Precision} & \textbf{Recall} & \textbf{F1-Score} \\ \hline
    
    \multirow{2}{*}{} Gemini & CySecBERT & 91.36\% & 90.07\% & 90.50\% & 91.25\% \\ 
    & SecBERT & 91.17\% & 90.73\% & 89.25\% & 91.04\% \\ \hline
    
    \multirow{2}{*}{} Llama & CySecBERT & 87.97\% & 86.10\% & 87.00\% & 87.85\% \\ 
    & SecBERT & 86.83\% & 86.46\% & 83.44\% & 86.62\% \\ \hline
    
    \multirow{2}{*}{} Mistral & CySecBERT & 88.89\% & 87.50\% & 87.50\% & 88.75\% \\ 
    & SecBERT & 86.00\% & 85.22\% & 82.88\% & 85.80\% \\ \hline
\end{tabular*}
\end{table*}

\subsubsection{Fusion Models Performance Comparison}
\begin{figure}
    \centering
\includegraphics[width=0.7\columnwidth]{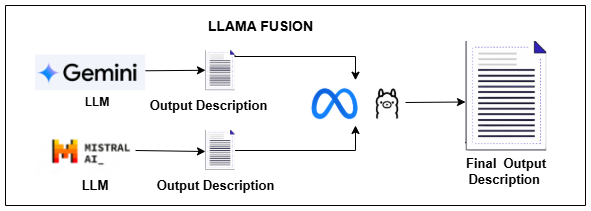}
    \caption{Architecture of Llama Fusion}
    \label{fig:llama-fusion}
\end{figure}

\begin{figure}
    \centering
\includegraphics[width=0.7\columnwidth]{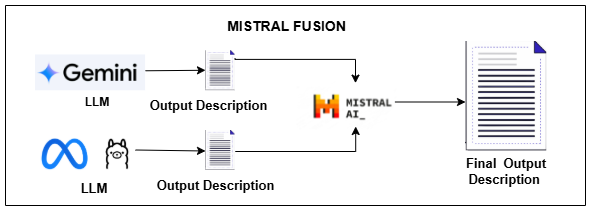}
    \caption{Architecture of Mistral Fusion}
    \label{fig:mistral-fusion}
\end{figure}

\begin{figure*}
  \centering
  \addcontentsline{lof}{section}{6.8.2} 
  
  \begin{subfigure}[b]{0.47\textwidth}
    \centering
    \includegraphics[width=\textwidth]{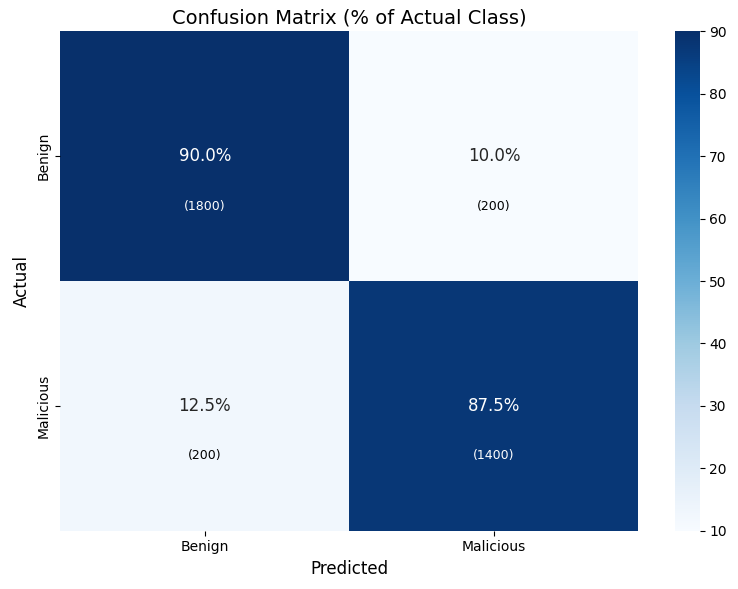}
    \caption{}
  \end{subfigure}
  \hfill
  \begin{subfigure}[b]{0.495\textwidth}
    \centering
    \includegraphics[width=\textwidth]{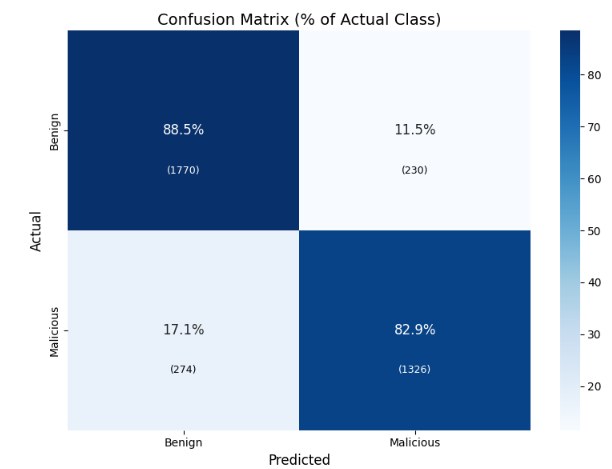}
    \caption{}
  \end{subfigure}

  \caption{Confusion Matrix for (a) Mistral Fusion Descriptions Classified using CySecBert, (b) Gemini Fusion Descriptions Classified using SecBert}
  \label{fig:CM3}
\end{figure*}

\begin{figure*}
  \centering
  \addcontentsline{lof}{section}{6.8.2} 
  
  \begin{subfigure}[b]{0.47\textwidth}
    \centering
    \includegraphics[width=\textwidth]{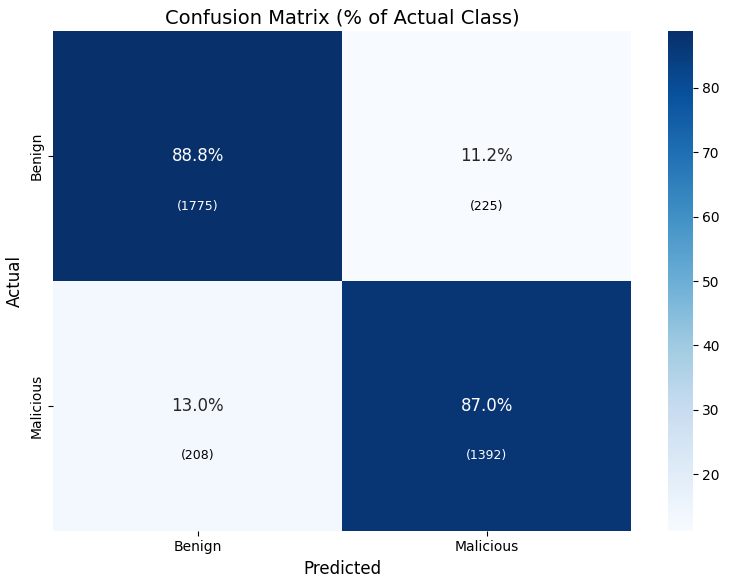}
    \caption{}
  \end{subfigure}
  \hfill
  \begin{subfigure}[b]{0.495\textwidth}
    \centering
    \includegraphics[width=\textwidth]{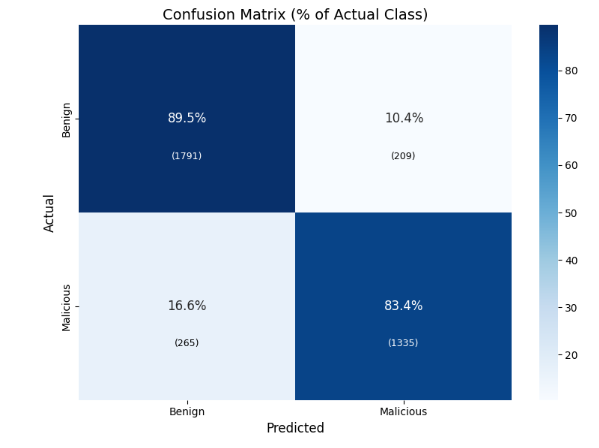}
    \caption{}
  \end{subfigure}

  \caption{Confusion Matrix for (a)Llama Fusion Descriptions Classified using CySecBert, (b)Gemini Fusion Descriptions Classified using SecBert}
  \label{fig:CM4}
\end{figure*}

As anticipated above, we conducted a comprehensive comparative analysis to identify the most effective large language model (LLM) for fusion tasks. Our evaluation systematically compared three distinct models: Gemini 2.0 Flash Lite, Llama2\footnote{\url{https://www.llama.com/llama2/}}, and Mistral\footnote{\url{https://mistral.ai/news/announcing-mistral-7b}}, each serving as the fusion component in our architecture.
The comparative framework employed a cross-validation approach where each model was evaluated as a fusion engine while receiving input descriptions from the other two models. Specifically, in the first configuration, descriptions generated independently by Gemini 2.0 Flash Lite and Mistral were provided as input to the Llama2 model for fusion processing (as illustrated in Figure \ref{fig:llama-fusion}). Conversely, in the second configuration, descriptions produced by Gemini 2.0 Flash Lite and Llama2 were fed into the Mistral model to perform the fusion operation (with the corresponding architecture detailed in Figure \ref{fig:mistral-fusion}). We have already seen Gemini as the fusion model previously in Figure \ref{fig:Gemini}. This systematic evaluation methodology enabled us to assess each model's fusion capabilities under consistent input conditions and determine the optimal LLM for our fusion pipeline.
All these descriptions were also pre-processed and classified using CySecBERT and SecBERT. The results shown in Table \ref{tab:model_comparison3} clearly demonstrate Gemini 2.0 Flash Lite as the best fusion model, outperforming Llama and Mistral across all four evaluation metrics. Gemini fusion descriptions, when classified using CySecBERT, obtained an accuracy of $91.36\%$, precision of $90.07\%$, recall of $90.50\%$, and F1-score of $91.25\%$. The confusion matrix for Llama fusion description classification is shown in Figure \ref{fig:CM3}, and the confusion matrix for Mistral fusion description classification is shown in Figure \ref{fig:CM4}.

The superior performance of AgenticRAG can be attributed to several factors:
\begin{enumerate}
    \item \textbf{Enhanced Contextual Specificity}: AgenticRAG descriptions captured application behaviors with greater technical precision, particularly in representing complex permission patterns and API call sequences that serve as strong malware indicators.
    \item \textbf{Improved Edge Case Handling: }The retrieval-augmented generation process demonstrated significantly better performance on applications exhibiting unusual permission combinations or obfuscated malicious behaviors—cases where Gemini Fusion often produced overgeneralized or imprecise characterizations.
    \item \textbf{Semantic Coherence: }The agentic planning component of AgenticRAG produced more coherent functional narratives that maintained logical consistency between different behavioral aspects of the analyzed applications.
    \item \textbf{Malware-Specific Terminology: }AgenticRAG descriptions incorporated more domain-specific security terminology, enabling the CySecBERT classifier to better leverage its pre-trained understanding of cybersecurity concepts.
\end{enumerate}
Although CySecBERT did not outperform SecBERT across every metric, it demonstrated superior recall and F1-score, making it more effective in identifying malicious instances. In the AgenticRAG setting, CySecBERT achieved a recall of $96.69\%$ and an F1-score of $92.86\%$, compared to SecBERT's lower values, indicating its stronger ability to correctly detect threats with fewer false negatives. Similarly, in the Gemini Fusion setting, CySecBERT maintained a better balance between precision and recall, resulting in a higher F1-score of $91.25\%$. These results justify the selection of CySecBERT for malware classification tasks, where maximizing threat detection (high recall) while maintaining consistent performance (high F1-score) is critical.\\
This advantage can be attributed to CySecBERT's more specialized pre-training on cybersecurity corpora and its enhanced ability to interpret malware-specific terminology and contextual relationships. The model's deeper domain knowledge enabled more accurate interpretation of technical patterns indicative of malicious intent, particularly in cases involving sophisticated obfuscation techniques common in modern Android malware.\\
These quantitative findings validate our hypothesis that integrating structured knowledge retrieval with agentic planning produces descriptions that not only enhance interpretability but also substantially improve classification performance in security-critical applications like Android malware detection. Furthermore, they confirm the importance of domain-specific model specialization in achieving optimal classification results.

%\clearpage
\section{Conclusion}
\label{sec:conclusion}
This research presents a sophisticated, multi-faceted approach to Android malware classification that integrates static feature extraction, AgenticRAG, and classification via domain-specialized language models. Through rigorous comparative analysis, the descriptions generated by AgenticRAG demonstrated superior performance compared to those produced by a Gemini-based fusion of LLaMA and Mistral architectures, exhibiting improved classification accuracy and semantic relevance.
The empirical findings substantiate the significant value of incorporating structured knowledge retrieval mechanisms, goal-directed generation processes, and agentic planning frameworks within the description generation pipeline. The implementation of CySecBERT, a model fine-tuned on cybersecurity-specific corpora, provided the classification framework with enhanced capabilities for interpreting complex and obfuscated malware behaviors that traditional approaches might overlook.
This methodological integration yielded consistent and statistically significant improvements in all the considered evaluation metrics, establishing AgenticRAG not merely as a superior description generation system but as a catalyst for more effective downstream malware detection. The research conclusively demonstrates the potential of agentic artificial intelligence systems, when strategically integrated with domain-specialized models and sophisticated retrieval methodologies, to address the increasingly complex challenges in security-critical applications.
These findings contribute substantially to the advancement of explainable and high-performance malware classification techniques, offering promising directions for future research in cybersecurity defense mechanisms and intelligent threat detection systems.

This paper can be considered a starting point for our research. 
Our future work will focus on three key directions to enhance our malware detection framework. First, we plan to incorporate additional static features to improve detection accuracy and capture a wider spectrum of malicious behaviors, including more sophisticated obfuscation techniques and semantic patterns. Second, we will expand AgenticRAG by integrating dynamic analysis capabilities, enabling the detection of evasive malware that activates only during runtime or employs dormancy techniques to avoid static analysis. Finally, we intend to extend our approach across multiple operating systems and application environments, including mobile platforms, to provide comprehensive security coverage against evolving threats in diverse computing ecosystems. These enhancements aim to address the current limitations of our framework while advancing the state-of-the-art in automated malware detection systems.

 \bibliographystyle{plain}
 \bibliography{biblio}

\begin{thebibliography}{10}

\bibitem{arzt2014flowdroid}
Steven Arzt, Siegfried Rasthofer, Christian Fritz, Eric Bodden, Alexandre Bartel, Jacques Klein, Yves Le~Traon, Damien Octeau, and Patrick McDaniel.
\newblock Flowdroid: Precise context, flow, field, object-sensitive and lifecycle-aware taint analysis for android apps.
\newblock {\em ACM sigplan notices}, 49(6):259--269, 2014.

\bibitem{felt2011android}
Adrienne~Porter Felt, Erika Chin, Steve Hanna, Dawn Song, and David Wagner.
\newblock Android permissions demystified.
\newblock In {\em Proceedings of the 18th ACM conference on Computer and communications security}, pages 627--638, 2011.

\bibitem{zhou2012dissecting}
Yajin Zhou and Xuxian Jiang.
\newblock Dissecting android malware: Characterization and evolution.
\newblock In {\em 2012 IEEE symposium on security and privacy}, pages 95--109. IEEE, 2012.

\bibitem{arp2014drebin}
Daniel Arp, Michael Spreitzenbarth, Malte Hubner, Hugo Gascon, Konrad Rieck, and CERT Siemens.
\newblock Drebin: Effective and explainable detection of android malware in your pocket.
\newblock In {\em Ndss}, volume 14(1), pages 23--26, 2014.

\bibitem{tam2015copperdroid}
Kimberly Tam, Aristide Fattori, Salahuddin Khan, and Lorenzo Cavallaro.
\newblock Copperdroid: Automatic reconstruction of android malware behaviors.
\newblock In {\em NDSS Symposium 2015}, pages 1--15, 2015.

\bibitem{meng2018droidecho}
Guozhu Meng, Ruitao Feng, Guangdong Bai, Kai Chen, and Yang Liu.
\newblock Droidecho: an in-depth dissection of malicious behaviors in android applications.
\newblock {\em Cybersecurity}, 1:1--17, 2018.

\bibitem{demontis2017yes}
Ambra Demontis, Marco Melis, Battista Biggio, Davide Maiorca, Daniel Arp, Konrad Rieck, Igino Corona, Giorgio Giacinto, and Fabio Roli.
\newblock Yes, machine learning can be more secure! a case study on android malware detection.
\newblock {\em IEEE transactions on dependable and secure computing}, 16(4):711--724, 2017.

\bibitem{zhang2025agentic}
Weinan Zhang, Junwei Liao, Ning Li, Kounianhua Du, and Jianghao Lin.
\newblock Agentic information retrieval, 2025.

\bibitem{suarez2017droidsieve}
Guillermo Suarez-Tangil, Santanu~Kumar Dash, Mansour Ahmadi, Johannes Kinder, Giorgio Giacinto, and Lorenzo Cavallaro.
\newblock Droidsieve: Fast and accurate classification of obfuscated android malware.
\newblock In {\em Proceedings of the seventh ACM on conference on data and application security and privacy}, pages 309--320, 2017.

\bibitem{onwuzurike2019mamadroid}
Lucky Onwuzurike, Enrico Mariconti, Panagiotis Andriotis, Emiliano~De Cristofaro, Gordon Ross, and Gianluca Stringhini.
\newblock Mamadroid: Detecting android malware by building markov chains of behavioral models (extended version).
\newblock {\em ACM Transactions on Privacy and Security (TOPS)}, 22(2):1--34, 2019.

\bibitem{rastogi2013droidchameleon}
Vaibhav Rastogi, Yan Chen, and Xuxian Jiang.
\newblock Droidchameleon: evaluating android anti-malware against transformation attacks.
\newblock In {\em Proceedings of the 8th ACM SIGSAC symposium on Information, computer and communications security}, pages 329--334, 2013.

\bibitem{suarez2013evolution}
Guillermo Suarez-Tangil, Juan~E Tapiador, Pedro Peris-Lopez, and Arturo Ribagorda.
\newblock Evolution, detection and analysis of malware for smart devices.
\newblock {\em IEEE communications surveys \& tutorials}, 16(2):961--987, 2013.

\bibitem{yan2012droidscope}
Lok~Kwong Yan and Heng Yin.
\newblock $\{$DroidScope$\}$: Seamlessly reconstructing the $\{$OS$\}$ and dalvik semantic views for dynamic android malware analysis.
\newblock In {\em 21st USENIX security symposium (USENIX security 12)}, pages 569--584, 2012.

\bibitem{wang2019effective}
Wei Wang, Mengxue Zhao, and Jigang Wang.
\newblock Effective android malware detection with a hybrid model based on deep autoencoder and convolutional neural network.
\newblock {\em Journal of Ambient Intelligence and Humanized Computing}, 10(8):3035--3043, 2019.

\bibitem{hou2017hindroid}
Shifu Hou, Yanfang Ye, Yangqiu Song, and Melih Abdulhayoglu.
\newblock Hindroid: An intelligent android malware detection system based on structured heterogeneous information network.
\newblock In {\em Proceedings of the 23rd ACM SIGKDD international conference on knowledge discovery and data mining}, pages 1507--1515, 2017.

\bibitem{liu2023summary}
Yiheng Liu, Tianle Han, Siyuan Ma, Jiayue Zhang, Yuanyuan Yang, Jiaming Tian, Hao He, Antong Li, Mengshen He, Zhengliang Liu, et~al.
\newblock Summary of chatgpt-related research and perspective towards the future of large language models.
\newblock {\em Meta-radiology}, 1(2):100017, 2023.

\bibitem{zhao2025apppoet}
Wenxiang Zhao, Juntao Wu, and Zhaoyi Meng.
\newblock Apppoet: Large language model based android malware detection via multi-view prompt engineering.
\newblock {\em Expert Systems with Applications}, 262:125546, 2025.

\bibitem{zheng2023judging}
Lianmin Zheng, Wei-Lin Chiang, Ying Sheng, Siyuan Zhuang, Zhanghao Wu, Yonghao Zhuang, Zi~Lin, Zhuohan Li, Dacheng Li, Eric Xing, et~al.
\newblock Judging llm-as-a-judge with mt-bench and chatbot arena.
\newblock {\em Advances in Neural Information Processing Systems}, 36:46595--46623, 2023.

\bibitem{sun2024research}
Ya~Sun, Kaiwen Yang, and Gong Chen.
\newblock The research trends of corpus-assisted stance research (2004--2023): a systematic literature review.
\newblock {\em Current Psychology}, pages 1--16, 2024.

\bibitem{bayer2024cysecbert}
Markus Bayer, Philipp Kuehn, Ramin Shanehsaz, and Christian Reuter.
\newblock Cysecbert: A domain-adapted language model for the cybersecurity domain.
\newblock {\em ACM Transactions on Privacy and Security}, 27(2):1--20, 2024.

\bibitem{huang2024secbert}
Hai Huang and Yongjian Wang.
\newblock Secbert: Privacy-preserving pre-training based neural network inference system.
\newblock {\em Neural Networks}, 172:106135, 2024.

\end{thebibliography}

\end{document}